\documentclass{article}

\usepackage{arxiv}
\pdfoutput=1
\usepackage[utf8]{inputenc} % allow utf-8 input
\usepackage[T1]{fontenc}    % use 8-bit T1 fonts

\usepackage{natbib}
\usepackage{hyperref}       % hyperlinks
\usepackage{url}            % simple URL typesetting
\usepackage{booktabs}       % professional-quality tables
\usepackage{amsfonts}       % blackboard math symbols
\usepackage{nicefrac}       % compact symbols for 1/2, etc.
\usepackage{microtype}      % microtypography
\usepackage{lipsum}		% Can be removed after putting your text content
\usepackage{amsmath}
\usepackage{graphicx}
\usepackage{epsfig}
\usepackage{doi}
\usepackage{tikz}
\usetikzlibrary{shapes.geometric, arrows}
\definecolor{darkgray}{rgb}{0.66, 0.66, 0.66}
\usepackage{rotating}
\usepackage{epstopdf}

\title{A Bayesian Approach to Modeling Variance of Intensive Longitudinal Biomarker Data as a Predictor of Health Outcomes}

\author{Mingyan Yu \\
	Department of Biostatistics\\
	University of Michigan\\
	Ann Arbor, MI 48109 \\
	\texttt{myanyu@umich.edu} \\
	\And
	{Zhenke Wu} \\
	Department of Biostatistics\\
	University of Michigan\\
	Ann Arbor, MI 48109 \\
	\And
    {Margaret Hicken} \\
	Institute for Social Research\\
	University of Michigan\\
	Ann Arbor, MI 48109 \\
    \And
    {Michael R. Elliott} \\
    Department of Biostatistics\\
	University of Michigan\\
	Ann Arbor, MI 48109 \\}
\date{}

\renewcommand{\shorttitle}{Joint Modeling for Intensive Longitudinal Predictors and Health Outcomes}

\hypersetup{
pdftitle={A template for the arxiv style},
pdfsubject={q-bio.NC, q-bio.QM},
pdfauthor={David S.~Hippocampus, Elias D.~Striatum},
pdfkeywords={First keyword, Second keyword, More},
}

\begin{document}
\maketitle

\begin{abstract}
	Intensive longitudinal biomarker data are increasingly common in scientific studies that seek temporally granular understanding of the role of behavioral and physiological factors in relation to outcomes of interest. Intensive longitudinal biomarker data, such as those obtained from wearable devices, are often obtained at a high frequency typically resulting in several hundred to thousand observations per individual measured over minutes, hours, or days. Often in longitudinal studies, the primary focus is on relating the means of biomarker trajectories to an outcome, and the variances are treated as nuisance parameters, although they may also be informative for the outcomes. In this paper, we propose a Bayesian hierarchical model to jointly model a cross-sectional outcome and the intensive longitudinal biomarkers. To model the variability of biomarkers and deal with the high intensity of data, we develop subject-level cubic B-splines and allow the sharing of information across individuals for both the residual variability and the random effects variability. Then different levels of variability are extracted and incorporated into an outcome submodel for inferential and predictive purposes. We demonstrate the utility of the proposed model via an application involving bio-monitoring of hertz-level heart rate information from a study on social stress.
\end{abstract}

% keywords can be removed

\section{Introduction}

Over the past several decades a large literature has developed to use
longitudinal measurements such as biomarkers or survey responses to predict health outcomes. For example, \citet{henderson2000joint} linked linear mixed effects models for longitudinal measures of psychiatric disorders with semi-parametric proportional hazard models for time-to-drop-out through a latent bivariate Gaussian process to explore the association between longitudinal psychiatric disorder scores and the outcome of interest. \citet{wu2008joint} extended the sub-model for longitudinal biomarkers into non-linear mixed effects models and connected them with the survival sub-model via subject-level random effects estimated from the longitudinal trajectories. \citet{horrocks2009prediction} used longitudinal measurements of the adhesiveness of CD56$^{\sf bright}$ cells to predict the probability of achieving a successful pregnancy. \\

\noindent

The development of ecological momentary assessment (EMA) data \citep{Shiffman2008}, which is typically provided at much higher frequencies with consequently much larger individual level sample sizes, has spawned a smaller but rapidly developing literature to incorporate EMA data into prediction of health outcomes. EMA data can occur in a variety of forms: for example, prompts on smartphone or other electronic devices to complete quesitonnaires or undertake tasks \citep{Schuster2016}; respondent tracking of events occurring in their experience \citep{Li2012}; or, as in our example, biometric measures such as skin conductivity or heart rates.  By virtue of its density, EMA data allows both assessment of short term effects and the possibility of more accurate prediction or association with longer-term outcomes. But this density as well as the variety of ways in which this data is collected also may require special consideration of statistical methods for it analysis \citep{Oleson2022}.  In general some type of method that accounts for the longitudinal nature of the data via ANOVA \citep{Schuster2016} or linear or generalized linear mixed models \citep{Li2012,Shiffman2020} or structural equation modeling \citep{McNeish2021} have been used. \\

\noindent

While a considerable proportion of studies relating longitudinal data to health outcomes have primarily focused on the mean structure of the longitudinal measurements and treated the variance as a nuisance parameter, it is crucial to realize that the variance structure itself can be of vital importance and can have a strong impact on the subsequent analysis \citep{carroll2003variances}. 
The literature in this area is small but growing; for example, \citet{palmier2012affective} studied the association between affective variability and suicidal ideation by regressing suicidal ideation scores on daily variability of negative and positive affect scores calculated by mean squared successive difference of the self-reported affect scores within each day. Mood variability was found to have a significant and stronger association with the frequency and severity of suicidality than the daily average.  \citet{taylor2016bedtime} concluded from the Study of Women's Health Across the Nation (SWAN) sleep study that greater variability in bedtime rather than later bedtime was associated with metabolic health in midlife women by relating the standard deviations of individuals’ sleep time over two weeks to insulin resistance. \citet{gao2011joint} found that individual-level variability of intraocular pressure is an independent risk factor for onset of primary open-angle glaucoma. \citet{elliott2012associations} used linked mean and variance of cognitive scores to development dementia, finding that variability (but not mean) was predictive of dementia onset. \citet{jiang2015joint}  considered a mixture model which included latent classes for the individual-specific mean profiles and residual variances, finding that higher variability in Follicle Stimulating Hormone (FSH) was associated with a higher chance of experiencing hot flashes in females. \\

\noindent

Consideration of variability as a direct predictor in EMA data does not seem to exist.  \citet{Hedeker2012} did model between- and within-variance of negative and positive affect obtained by random prompts from handheld computers as a function of baseline covaraites via long-linear models, but did not consider variability of the affect measures themselves as predictors. \\

\noindent

A key feature of EMA data is that it is obtained at a high frequency and typically results in hundreds to thousands of observations per individual taken over hours or days.  Furthermore, here we focus on heart rate, which due to autonomic regulation of the cardiovascular system is approximately stationary except in very short time intervals, so that mean trends are not available as predictors.  Heart rate variability (HRV) has been considered as a predictor of several health outcomes outside of EMA settings \citep{Kikuya2008,Hsu2016}. 
The time and frequency domain of HRV are first derived based on the interbeat intervals, which are then averaged over a specific time period (e.g., five-minute average heart rate variability). A previous study on HRV and stress have shown that standard deviations of interbeat intervals (SDNN), which measures the overall HRV, is significantly lower when people were under stress \citep{endukuru2016evaluation}. \\

\noindent

Instead of simple averaging, our approach to analysis of heart rate data and HRV is different and, we believe, highly novel and likely to access quite a bit more infomation from standard HRV.  First, we model variability in heart rates using a random effects B-spline seprate variance into a longer-term variability component measured by the spline random effects variance (``wiggliness'' of the heart rate) and a shorter-term variability component measured by the residual variance (``second-to-second'' variability of the heart rate).  We also consider an individual autocorrelation measure \citep{Du2018} to account for correlation the in residual variance and to serve as a possible preditor as well. We use a joint model, with the individual level ``wiggleness'' and ``second-to-second'' variability together with residual autocorrelation modeling assignment to one of four controlled stress settings.  This joint modeling is a key feature of our development as well, since two-stage approaches that estimate variabilities of longitudinal measurements  in the first stage and then plug in these sample-based estimates into the second stage outcome model will usually lead to biased inference and uncontrolled variance in the second stage model \citep{wang2020methods} as the uncertainty of the mean and variance estimation in the first stage is not properly propagated to the outcome model. 
Corrections have been proposed to eliminate the bias in two-stage models.  For example, \citet{ye2008semiparametric} proposed a risk set regression calibration (RRC) method to correct the bias inherent in the naive two-stage model when connecting longitudinal measurements and the time-to-event outcome of interest; however, their method worked well only when measurement error in the longitudinal marker was low. Joint modeling approaches have attracted increasing attention in the past two decades and simulation studies have shown that joint modeling the longitudinal measurements and cross-sectional or survival outcomes together generally leads to unbiased results \citep{wu2012analysis}. These joint models are often implemented using shared random effects models \citep{wu1998}, where a first stage model is used to estimate subject-level trends using random effects, and these random effects are then used as predictors in the second stage models \citep{Ibrahim2010, Alsefri2020}.  We follow such an approach here. \\

\noindent

We demonstrate the application of our work to the Michigan Work-Life Study where we aim to determine how variability in hertz-level heart rate measurements are associated with stress, which is reflected by the group assignments. This study was designed to examine the nature of race-based anticipatory and perseverative stress among Black women. Using the Trier Social Stress Test (TSST) to elicit a social stress response (analogous to the use of the treadmill to elicit a physical stress response), participants were randomly assigned to one of four race-related stress exposure groups (described in detail below). The TSST includes three phases: (a) anticipation, which captures anticipatory stress; (b) social stress, which is the actual acute stress exposure; and (c) recovery, which may capture any perseverative stress. The biological stress responses during the TSST can be measured in numerous ways; in the Michigan Work-Life Study, these responses were measured using blood pressure, electrodermal activity, heart rate, salivary cortisol, and salivary alpha-amylase. In this applied demonstration of this novel statistical approach, we focus on hertz-level heart rate information. 
We take a non-standard approach in this analysis, in part for illustrative purposes of the method, treating the assignment group as the outcome. Thus our analysis will focus on understanding how variability in heart rate is associated with the various treatment assignments rather than viewing variability strictly as a predictor in a formal sense. \\

\noindent

The remainder of our manuscript is organized as follows.  Section 2 develops the statistical details of the proposed joint model in a Bayesian framework.  Section 3 implements a simulation study to assess the quality of inference from the joint model, and compares this with somewhat simpler a two-stage approach.  Section 4 applies our method to the Michigan Work-Life Study.  Section 5 summarizes the findings of the simulation study and application and considers limitations and the next step for research.

\section{METHODS}

\subsection{Observed data}

Let $Y_i$ denote the cross-sectional binary outcome of subject $i = 1,...N$. Let $X_{ij}$ denote the value of the longitudinal biomarker of subject $i$ at 
%equally spaced
observational time points $t_{ij}$, $j = 1,...,n_i$. 
%which may vary across participants.
Baseline covariates are denoted by $\boldsymbol{Z}_i$.

\subsection{Joint Model}

To jointly model the longitudinal biomarker and the binary outcome together, the proposed model is comprised of two components. In the longitudinal marker sub-model, we model the intensive longitudinal marker data $X_{ij}, i=1, \ldots, N, j=1, \ldots, n_i$ as a non-linear function of $t_{ij}$
using a B-spline regression model.
The outcome sub-model then links the  
short-term and long-term
variances from the B-spline 
longitudinal marker sub-model with the binary outcome $Y_i$.

\subsubsection{Predictor Sub-model: Longitudinal Biomarker}

The proposed model for the longitudinal marker is as follows:
\begin{align}
    & \boldsymbol{X_i}\sim \boldsymbol{N}_{n_i}(f(\boldsymbol{t_i}; \boldsymbol{\beta}, \boldsymbol{b_i}), \boldsymbol{\Sigma_i}), \text{independently for } i = 1,...,N, \nonumber \\
    & b_{il}\overset{\mathrm{i.i.d}}\sim N(0, \sigma_{b_i}^2), l = 1,...,L.
\end{align}
Here, $\boldsymbol{N}_{n_i}(\boldsymbol{f}_i, \boldsymbol{\Sigma_i})$ denotes an $n_i$-dimensional multivariate normal distribution with mean $\boldsymbol{f}_i$ and covariance matrix $\boldsymbol{\Sigma_i}$, where $n_i$ represents the number of observations for participant $i$. The function $f(\boldsymbol{t_i}; \boldsymbol{\beta}, \boldsymbol{b_i})$ represents a 
%linear or 
non-linear function of time $\boldsymbol{t_i}$ characterized by population-level fixed effects $\boldsymbol{\beta}$ and individual-level random effects $\boldsymbol{b_i}$. In our case, this non-linear function is a B-spline regression model provided by $f(t_{ij};\boldsymbol{\beta},\boldsymbol{b}_i)=\beta_0+\sum_{l=1}^Lb_{il}\Phi_{l,d}(t_{ij})$. The individual-level random effects $\boldsymbol{b_i}=(b_{i1},...,b_{iL})$ is a vector of length $L$ where $L$ denotes the number of basis functions of time and we assume participants have the same number of basis functions here. $\Phi_{l,d}(\cdot)$ is the B-spline basis function of degree $d$ and is formulated by $\Phi_{l,d}(t_{ij})=\frac{t_{ij}-k_l}{k_{l+d}-k_l}\Phi_{l,d-1}(t_{ij})+\frac{k_{l+1+d}-t_{ij}}{k_{l+1+d}-k_{l+1}}\Phi_{l+1,d-1}(t_{ij})$ with $\Phi_{l,0}(t_{ij})=I(k_l\leq t_{ij}\leq k_{l+1})$; $k_l$'s in the formulation are the time knots based on specific selections. In our simulation and application, we set the degree of the B-spline basis function to be 3 and placed a time knot at every 2nd percentile of the observed time duration. Since intensive longitudinal measurements are likely to be correlated with each other, we can impose different covariance matrices to reflect the correlations. We can rewrite $\boldsymbol{\Sigma}_i$ as $\sigma_{e_i}^2\boldsymbol{R}_i$, where we assume that observations for participant $i$ have a constant residual variance $\sigma_{e_i}^2$ across time and $\boldsymbol{R}_i$ is the correlation matrix.\\

\noindent
This model uses individual-specific variability in the random effects, $\sigma_{b_i}^2$'s, to characterize ``long-term" variability (i.e., minute-to-minute variability) in the longitudinal marker, and 
uses individual-specific variability in the residual errors, $\sigma_{e_i}^2$'s, to characterize ``short-term" variability (i.e., second-to-second variability) in the longitudinal marker. The traditional definition of ``Heart Rate Variability" is the variation in the duration between consecutive heart beats, also called an ``interbeat interval", as the heart continuously adapts to the environment, and includes numerous time- and frequency-based components important in clinical settings \citep{shaffer2017overview}. Faster heart rates mean that the interbeat interval is shorter, with less opportunity for variation; lower heart rates mean that the interbeat interval is longer, allowing for more opportunity for variation in heart rate or greater heart rate variability. Basic time-based measures of heart rate variability are generally calculated as the standard deviation or the root mean square of successive interbeat intervals over a short epoch, yielding single average values of heart rate variability. However, here, we examine a different approach to quantifying this time-based variability.

\subsubsection{Outcome Sub-model: Binary Outcome}

The binary outcome $Y_i$ is assumed to be associated with the individual-specific variances and other baseline covariates $\boldsymbol{Z}_i$ through the following probit regression model:
\begin{align}
    & S_i = [\sigma_{b_i}, \sigma_{e_i}, \boldsymbol{Z}_i]^T, \nonumber \\
    & Y_i\sim \text{Bern}(\Phi(\eta(S_i))), \nonumber \\
    & \eta(S_i)=\alpha_1+\alpha_2\times \log(\sigma_{b_i})+\alpha_3\times \log(\sigma_{e_i})+\boldsymbol{\gamma}\boldsymbol{Z}_i,  i = 1,...,N.
\end{align}
$S_i$ is a vector of individual-specific short-term variability $\sigma_{e_i}$, long-term variability $\sigma_{b_i}$ and baseline covariates $\boldsymbol{Z}_i$. The outcome is modeled using a logistic regression with a probit link; $\Phi$ in (2) denotes the cumulative density function (CDF) of the standard normal distribution. $\eta(S_i)$ is a linear function of log-transformed variances and baseline covariates. (In our example we do not include baseline covariates since the randomization of the $Y$ means that there was little imbalance with respect to covariates, but we include them in our model for completeness.) Although we specify a probit model for a binary outcome, a generalized linear model with a specified link function can be considered.

\subsection{Priors} 
 
We specify the following
priors and hyperpriors for each unknown parameter. \\

\noindent
For the longitudinal marker sub-model fixed effects $\boldsymbol{\beta}=(\beta_1,...,\beta_Q)$, we assume 
\begin{align}
    &\beta_q \sim N(0, \psi^2),
\end{align}
independently for $q=1,...,Q$. \\

\noindent
To allow sharing of information across individuals for long-term variability, we propose a common prior distribution for $\sigma_{b_i}$'s 
\begin{align}
    \log(\sigma_{b_i}^2)\sim N(v_b, \Psi_b^2),
\end{align}
with hyper-parameter priors \citep{Gelman2006}
\begin{align}
    v_b\sim N(m_b, \kappa_b^2), \Psi_b\sim \text{Half-Cauchy}(0, \tau_b).
\end{align}

\noindent
A general form for the prior distribution for the subject-level correlation matrix $\boldsymbol{R}_i$ is difficult to specify for $n_i>2$, given the need for $\boldsymbol{R}_i$ to be positive definite. Options to deal with this include positing independent $U(-1,1)$ distribution for correlation parameters and rejecting those that fail to be jointly positive definite \citep{Conlon2014} and parameterization of correlations in terms
of hyperspherical coordinates \citep{Ghosh2021}. Here we elect to take advantage of a high degree of beat-to-beat autocorrelation and assume an AR1 model so that the off-diagonal elements of the correlation matrix $\boldsymbol{R}_i$ become $r_{i,ks}=\rho_i^{|s-k|}$ for $1 \leq k,s \leq n_i$ with $\rho_i$ representing the autocorrelation parameter for individual $i$. We implement this model by decomposing the original error $\epsilon_{ij}=\rho_i\epsilon_{ij-1}+w_{ij}$ into a deterministic function of the error at the previous time point, $\rho_i\epsilon_{ij-1}$, and an innovative new ``white noise" $w_{ij}$, 
where $w_{ij}\overset{\mathrm{i.i.d}}\sim N(0, \sigma_{w_i}^2)$.
We assume the priors for this as
\begin{align}
    & \log(\sigma_{w_i}^2)\sim N(v_w, \Psi_w^2) \nonumber \\
     & v_w\sim N(m_w, \kappa_w^2), \Psi_w\sim \text{Half-Cauchy}(0, \tau_w) \nonumber \\ 
     & \rho_i \sim U(0,1)
\end{align}
where the prior on $\rho_i$ assumes a positive autocorrelation that is biologically warranted for heart rates. \\

\noindent
For the outcome sub-model, we assume an independent $N(0, \omega^2)$ priors for each of the regression coefficients $(\alpha_1,\alpha_2,\alpha_3,\boldsymbol{\gamma})$. \\

\noindent
For the hyperparameters, we chose values that correspond to weakly informative values.  Thus for the fixed effect regression hyperparameters, we set $\psi=1000$ and for the outcome sub-model regression hyperparameters, we set $\omega=10$.  For the variance hyperparamter model, we set $m_b=m_w=0$, $\kappa_b=\kappa_w=1000$, and $\tau_b=\tau_w=2.5$.

\subsection{Joint Distribution}

Let $\Theta=(\boldsymbol{\beta}, v_b, \Psi_b, v_w, \Psi_w, \rho_{i}, \alpha_1, \alpha_2, \alpha_3, \boldsymbol{\gamma})$ denotes the vector of unknown parameters of interest in the joint model. From the prior specification, we have assumed that the prior distributions for the unknown parameters are all independent. Let $\pi(\Theta)$ denote the prior distribution of the unknowns, then $\pi(\Theta)=\pi(\boldsymbol{\beta})\pi(v_b)\pi(\Psi_b)\pi(v_w)\pi(\Psi_w)\prod_{i=1}^N\pi(\rho_{i})\pi(\alpha_1,\alpha_2,\alpha_3)\pi(\boldsymbol{\gamma})$, the joint distribution of the data and unknowns is
\begin{align}
    & P(\boldsymbol{Y}, \boldsymbol{X}, \boldsymbol{t}, \Theta) \propto \prod_{i=1}^N \frac{1}{\sqrt{(2\pi)^{n_i}|\boldsymbol{\Sigma}_i|}}\exp\Bigl(-\frac{1}{2}[\boldsymbol{X}_i-f(\boldsymbol{t}_i;\boldsymbol{\beta}, \boldsymbol{b}_i)]^T\boldsymbol{\Sigma}_i^{-1}[\boldsymbol{X}_i-f(\boldsymbol{t}_i;\boldsymbol{\beta}, \boldsymbol{b}_i)]\Bigl) \nonumber \\
    & \times \prod_{l=1}^L\Bigl\{\frac{1}{\sqrt{2\pi\sigma_{b_i}^2}}\exp\Bigl(-\frac{b_{il}^2}{2\sigma_{b_i}^2}\Bigl)\Bigl\}\times  \frac{1}{\sqrt{2\pi\Psi_b^2}}\exp\Bigl\{-\frac{[\log(\sigma_{b_i}^2)-v_b]^2}{2\Psi_b^2}\Bigl\} \nonumber \\
    & \times \frac{1}{\sqrt{2\pi\Psi_w^2}}\exp\Bigl\{-\frac{[\log(\sigma_{w_i}^2)-v_w]^2}{2\Psi_w^2}\Bigl\} \times \Phi(\eta(\sigma_{b_i},\sigma_{w_i},\boldsymbol{Z}_i))^{Y_i}\times (1-\Phi(\eta(\sigma_{b_i},\sigma_{w_i},\boldsymbol{Z}_i)))^{1-Y_i} 
    \times \pi(\Theta).
\end{align}

\subsection{Posterior Inference}

Since we proposed several non-conjugate priors in our model, the 
posterior distribution is not available in closed form. Therefore, we implemented our joint model using \textbf{JAGS} and the \textbf{R2jags} package in R \citep{R2jags}. \textbf{JAGS} has several built-in sampling algorithms for non-conjugate distributions, including slice sampling, adaptive random walk Metropolis algorithm and etc. The main engine for \textbf{JAGS} is slice sampling, but \textbf{JAGS} can automatically choose samplers for updating parameter values. One benefit of choosing \textbf{JAGS} to implement our model is that the built-in glm module in \textbf{JAGS} is able to do block updating of $\boldsymbol{\beta}$ and $\boldsymbol{b}_i$ as it recognizes the generalized linear mixed model structure in the longitudinal marker sub-model, which substantially increases the sampling efficiency.

\section{Simulation Study}

Before applying our proposed model to any application settings, we conducted simulation studies. There are two goals associated with the simulation: (1) To test the performance of our proposed model to see whether it has the capacity to recover the true parameter values in terms of bias in the posterior means, posterior variance, empirical coverage rates and average lengths of the nominal $95\%$ credible intervals; (2) To compare with alternative two-stage approaches that can be used to obtain variances and then use variances as predictors of outcomes. Through our simulations, we demonstrated that our proposed joint model maintained satisfactory performance regardless of the magnitude of the measurement error in the longitudinal marker, whereas the commonly used two-stage approach failed to provide good results as the measurement error increased. Let $\theta_0$ denote the true parameter value of $\theta$ and $\hat{\theta}_r$ denote the posterior mean obtained from the r-th simulation. We evaluated the performance of the simulations based on the following four criteria: (1) \textbf{Coverage Rate} by calculating $\frac{1}{R}\sum_{r=1}^RI(\theta_0\in [L_r, U_r])$ where $L_r$ and $U_r$ represents the 2.5\% and 97.5\% percentiles of the posterior draws in the r-th simulation; (2) \textbf{Bias} by calculating $\frac{1}{R}\sum_{r=1}^R(\hat{\theta}_r-\theta_0)$; (3) \textbf{Average Interval Length} by calculating $\frac{1}{R}\sum_{r=1}^R(U_r-L_r)$. (4) \textbf{Root Mean Square Error (RMSE)} by calculating $\sqrt{\frac{1}{R}\sum_{r=1}^R(\theta_0-\hat{\theta}_r)^2}$.

For the simulations, we ran 3 chains and for each chain we ran 8,000 iterations with the first 4,000 as burn-in. We assessed the convergence of the models by Gelman-Rubin $\hat{R}$ measure. $\hat{R}$ is approximately the square root of the ratio of the total variance of a given set of parameter draws to their within-chain variance; the $\hat{R}$ values for all model parameters are smaller than 1.1, which is generally viewed as a satisfactory threshold \citep{gelman2013bayesian}.

\subsection{Simulation Settings}\label{subsec1}

In the simulation, we generated $n_i=600$ observations for each individual with $N=150$ individuals in each simulation, and we conducted $R=200$ replicates of the simulation. To mimic the heart rate data that would be modeled from the Michigan Work-Life Study in the Application section, we used the heart rate data in the Application section to obtain the true parameter values of the longitudinal marker sub-model that would be applied to our simulations. Specifically, we fit the longitudinal marker sub-model to 40 randomly selected participants' heart rate data from the Application data set and we used the exact same priors as described in the previous section. Note that we selected the time knots $k_l$'s in the cubic B-splines to be every 2nd percentile of the observed time points. As we mentioned previously, ignoring the uncertainty in the first stage, which partially arises from the measurement errors in the longitudinal data, can lead to biased inference in two-stage approaches. Therefore, we conducted two sets of simulation comparisons with different magnitudes of measurement errors in the longitudinal marker by varying the prior distribution of $\sigma_{w_i}$'s. Specifically, in the low measurement error simulation setting, we drew $\sigma_{w_i}$'s from a log-normal distribution with a mean $v_w$ and variance $\Psi_w^2$, where $v_w$ and $\Psi_w$ were set to the values obtained from fitting the application data mentioned above. In the high measurement error simulation setting, we drew $\sigma_{w_i}'s$ from the exact same distribution but we increased $v_w$ to a larger value.

\subsubsection{Simulation Setting 1: Low measurement error in the longitudinal marker}

\begin{figure}
    \centering
    \includegraphics[scale=0.75]{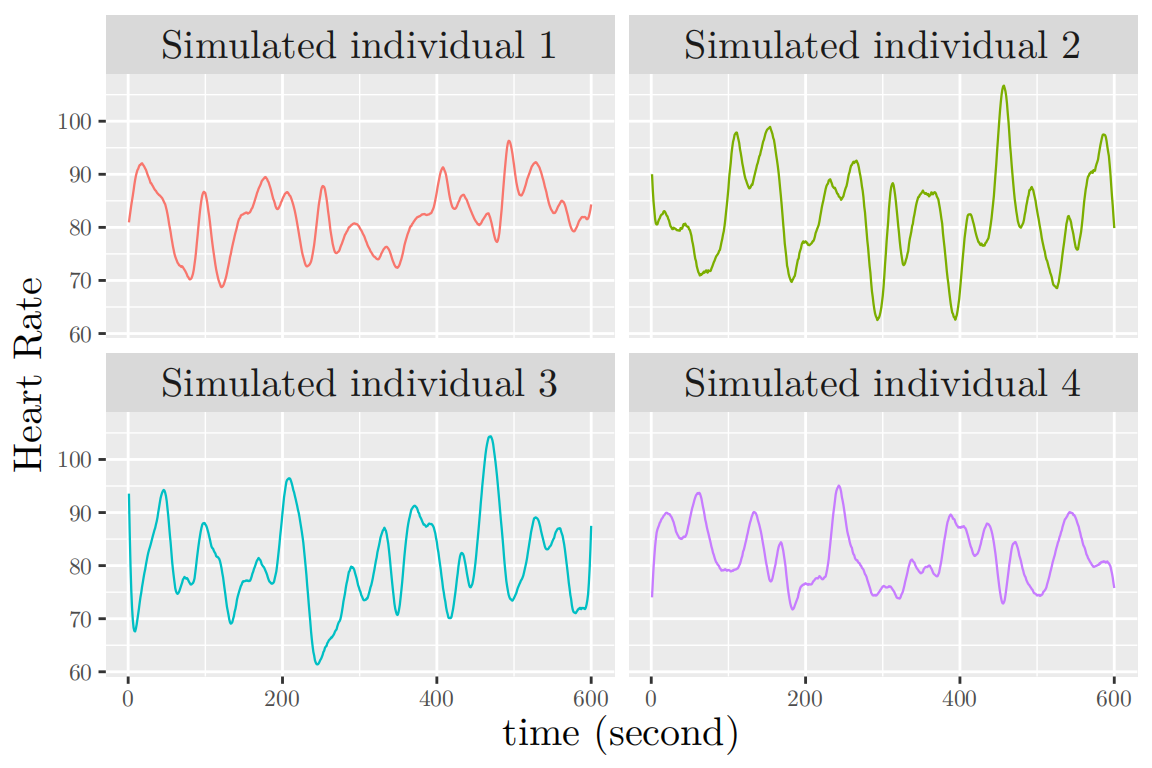}
    \caption{Simulated heart rate data of four randomly selected individuals in simulation setting 1.}
    \label{fig:simulatedHR}
\end{figure}

We first simulated the longitudinal marker data for each individual independently based on the parameter values we obtained from the application data set:
\begin{align}
    & X_{ij}=\beta_0+\sum_{l=1}^Lb_{il}\Phi_{l,3}(t_{ij})+\epsilon_{ij}, \ \ j=1,...,600; \ \ b_{il}\sim N(0, \sigma_{b_i}^2), \ \ l=1,...,L, \nonumber \\
    & \epsilon_{ij}=\rho_i\epsilon_{ij-1}+w_{ij},\ \ w_{ij}\overset{\mathrm{i.i.d}}\sim N(0, \sigma_{w_i}^2), \ \ j=1,...,600, \nonumber \\
    & \beta_0=81.083,\ \ \log(\sigma_{b_i}^2)\sim N(5, 1^2),\ \ \log(\sigma_{w_i}^2)\sim N(\boldsymbol{-3.8}, 0.75^2),\ \ \rho_i=0.998.
\end{align}
A sample of simulated heart rate data of four randomly selected individuals in one simulation is shown in Figure \ref{fig:simulatedHR}. From the figure, we notice that the simulated values fall into a reasonable range of heart rate and the trend looks like normal heart rate data, which means that our proposed longitudinal marker sub-model is able to generate HR-like data from reasonable parameter values.

As the next step, we generated the binary outcomes based on $\sigma_{b_i}$'s and $\sigma_{w_i}$'s simulated from the longitudinal sub-model. We assumed that the outcome $Y_i$'s came from a Bernoulli distribution with $P(Y_i=1)=\Phi(\eta(S_i))$, where $\Phi$ is the CDF of distribution N(0,1) and set
\begin{align}
    \eta(S_i)=\alpha_1+\alpha_2\times \log(\sigma_{w_i}\times 10)+\alpha_3\times \log\Bigl(\frac{\sigma_{b_i}}{10}\Bigl).
\end{align}
We scaled $\sigma_{b_i}$'s and $\sigma_{w_i}$'s inside the log function to facilitate model convergence. We set the true values of $\boldsymbol{\alpha}$ to be (0.3, -1, 0.5) so that the generated binary outcomes have a similar number of cases and controls.

\subsubsection{Simulation Setting 2: High measurement error in the longitudinal marker}

\begin{figure}
    \centering
    \includegraphics[scale=0.75]{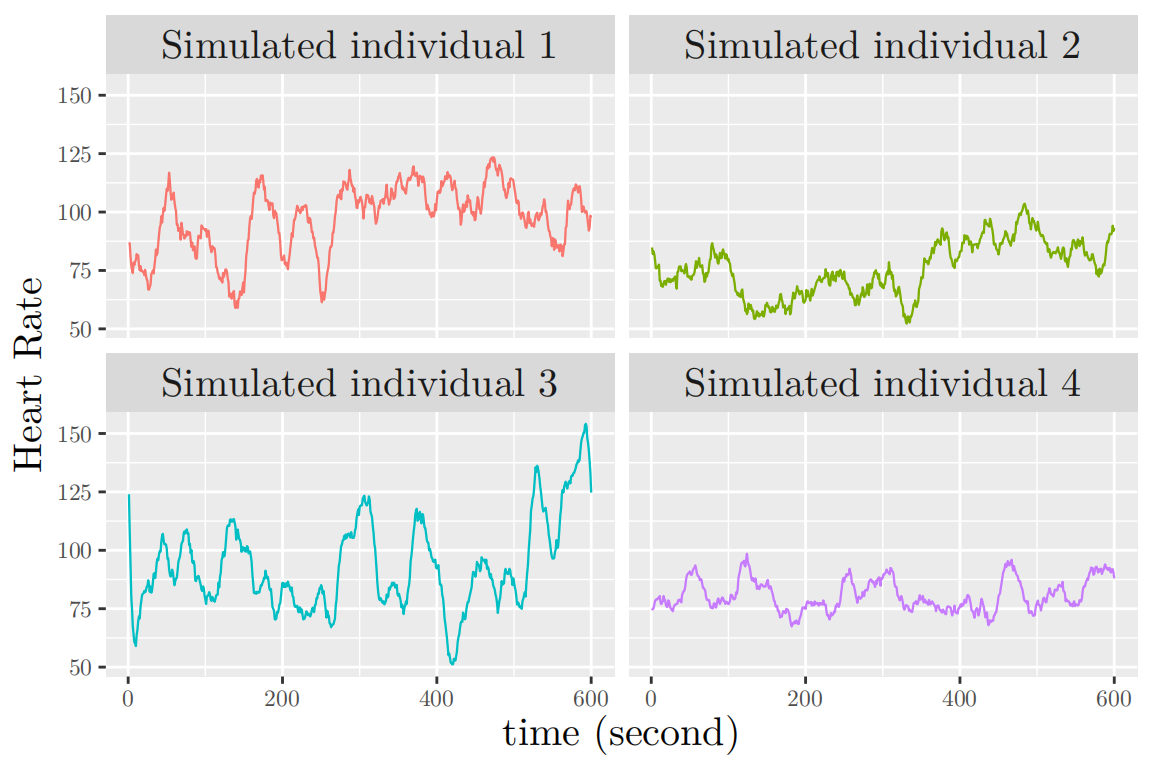}
    \caption{Simulated heart rate data of four randomly selected individuals in simulation setting 2.}
    \label{fig:simulatedHR2}
\end{figure}

In the second set of simulations, we chose a larger value for $v_w$, mean of the log-normal distribution from which $\sigma_{w_i}$'s were drawn. Other than that, all other parameter values in the longitudinal marker sub-model remained the same as in simulation setting 1:
\begin{align}
    & X_{ij}=\beta_0+\sum_{l=1}^Lb_{il}\Phi_{l,3}(t_{ij})+\epsilon_{ij}, \ \  j=1,...,600; \ \ b_{il}\overset{\mathrm{i.i.d}}\sim N(0, \sigma_{b_i}^2), \ \ l=1,...,L, \nonumber \\
    & \epsilon_{ij}=\rho_i\epsilon_{ij-1}+w_{ij},\ \ w_{ij}\overset{\mathrm{i.i.d}}\sim N(0, \sigma_{w_i}^2), \ \ j=1,...,600, \nonumber \\
    & \beta_0=81.083,\ \ \log(\sigma_{b_i}^2)\sim N(5, 1^2),\ \ \log(\sigma_{w_i}^2)\sim N(\boldsymbol{1.8}, 0.75^2),\ \ \rho_i=0.998.
\end{align}
Figure \ref{fig:simulatedHR2} displays the simulated heart rate data of four randomly selected individuals under the high measurement error setting. We notice that the simulated values and trends still resemble real heart rate data except that there are more fluctuations and the overall trends appear to be less smooth than the simulated data under the low measurement error setting.

As the next step, we generated the binary outcomes based on the two types of variances simulated from the longitudinal sub-model. We used the same procedure for drawing the outcome $Y_i$'s; however, we adjusted how we scaled $\sigma_{b_i}$'s and $\sigma_{w_i}$'s inside the log function to facilitate model convergence and we assigned new values to the coefficients $\boldsymbol{\alpha}$ to ensure the generated binary outcomes were balanced. Specifically, we set
\begin{align}
    \eta(S_i)=\alpha_1+\alpha_2\times \log(\sigma_{w_i})+\alpha_3\times \log\Bigl(\frac{\sigma_{b_i}}{10}\Bigl).
\end{align}
and we set the true values of $\boldsymbol{\alpha}$ to be (-0.6, 1, -1.5), again retaining an approximately equal number of positive and negative outcomes.

\subsection{Alternative Methods}

To compare with our joint model, we used a two-stage modeling strategy as an alternative method. In the first stage, we fit a linear mixed effects model for the simulated heart rate data across all simulated participants using lme() function from the \textbf{nlme} package in R:
\begin{align}
    X_{ij} = \beta_0 + \sum^L_{l=1}b_{il}\Phi_{l,3}(t_{ij})+\epsilon_{ij}.
\end{align}
When fitting the linear mixed effects model, we assumed that the residuals had a constant variance across time and followed a first-order autoregressive (AR1) structure.

To obtain the subject-specific long-term variability $\hat{\sigma}^2_{b_i}$'s, for each subject $i$, we calculated the empirical variance of its estimated random effects:  $\hat{\sigma}^2_{b_i}=\frac{1}{L-1}\sum_{l=1}^L (\hat{b}_{il}-\bar{\hat{b}}_i)^2$, $\bar{\hat{b}}_i=\frac{1}{L} \sum_{l=1}^L \hat{b}_{il}$.
%$\hat{b}_{i1},...,\hat{b}_{iL}$. 
To approximate the subject-specific short-term variability $\hat{\sigma}_{w_i}$'s, for each subject $i$, we first calculated the fitted values $\hat{X}_{ij}=\hat{\beta}_0+\sum^L_{l=1}\hat{b}_{il}\Phi_{l,3}(t_{ij})+\hat{\rho}\hat{\epsilon}_{ij-1}$ to get the estimated ``white noise" $\hat{w}_{ij}=X_{ij}-\hat{X}_{ij}$ and then computed $\hat{\sigma}^2_{w_i}=\frac{1}{n_i-1}\sum_{j=1}^{n_i} (\hat{w}_{ij}-\bar{\hat{w}}_i)^2$, $\bar{\hat{w}}_i=\frac{1}{n_i} \sum_{j=1}^{n_i} \hat{w}_{ij}$.

In the second stage, we regressed the simulated outcome on the scaled long-term variability $\log(\frac{\hat{\sigma}_{bi}}{10})$ and short-term variability $\log(\hat{\sigma}_{w_i}\times 10)$ or $\log(\hat{\sigma_{w_i}})$ using a logistic regression with a probit link function by glm() function in R.

\subsection{Simulation Results}

Table \ref{tab:compare-low} shows the results of comparing the two-stage approach with our proposed joint model when the measurement error in the longitudinal marker is low. For our proposed joint model, we observe that the coverage rates of the true parameter values in the outcome sub-model are all greater than or equal to 94\% and the estimated coefficients are nearly unbiased. The two-stage model has a slightly larger bias and somewhat lower coverage in this low-variance setting.
%The two-stage model also has a good performance with coverage rates all exceeding 90\%. The joint model exhibits a uniformly higher coverage rate; however, the differences are minimal. Moreover, both approaches yield similar interval lengths and similar amount of bias. Table \ref{tab:simulation_other_parms_low} displays the simulation results for all other parameters from the joint model. The results demonstrate that the joint model achieved high coverage rates and very low bias. These results provide evidence that our proposed model has the capacity to estimate parameters accurately and produce unbiased estimates under the low measurement error setting.

\begin{table}
    \begin{center}
    \caption{Simulation results including coverage rate, bias, average interval length and RMSE of $\boldsymbol{\alpha}$ comparing joint model and two-stage approach under simulation setting 1 when the measurement error in the longitudinal marker is low.\label{tab:compare-low}} %
    \begin{tabular}{@{}llllll@{}}
    \toprule
    True Values & Model & Coverage (\%) & Average Interval Length & Bias & RMSE\\
    \midrule
    $\alpha_1$ = 0.3 & Joint Model & 96.5 & 0.65 & 0.01 & 0.15\\ 
      &  Two-Stage Model & 94.0 & 0.63 & -0.05 & 0.17\\
    $\alpha_2$ = -1 & Joint Model & 95.0 & 1.20 & -0.02 & 0.28\\
      & Two-Stage Model & 93.0 & 1.22 & 0.01 & 0.34\\
    $\alpha_3$ = 0.5 & Joint Model & 94.0 & 0.91 & 0.01 & 0.21\\
      & Two-Stage Model & 92.0 & 0.87 & 0.03 & 0.24\\
    \bottomrule
    \end{tabular}
    \end{center}
\end{table}

Table \ref{tab:compare-high} presents the results of comparing the two approaches when the measurement error in the longitudinal marker is high. We observe that our proposed joint model still retained satisfactory bias and coverage rates of parameters in the outcome sub-model, 
although the interval lengths are wider than in the low variance setting. However, the two-stage approach now suffers from severe bias and low coverage rates due to the measurement error introduced through use of empirical estimates in the two-stage modeling.
%whereas the coverage rates from the two-stage approach are all noticeably lower. Although the average interval lengths and bias from the joint model are higher compared to the low measurement error setting, the estimated parameters are still nearly unbiased and the joint model outperforms the two-stage approach in terms of all evaluation metrics. Table \ref{tab:simulation_other_parms_high} presents the simulation results for the remaining parameters estimated by the joint model. Similar to simulation setting 1, the joint model results in high coverage rates and very low bias. These findings provide support that our proposed joint model can still produce accurate and unbiased estimates under the high measurement error setting. In contrast, the two-stage approach demonstrates its limitations and fail to achieve satisfactory results.

The main take-away from the two simulation comparisons is that while the two-stage approach can yield comparable results to our proposed joint model when the measurement error in the longitudinal marker is low, the joint model is superior over the two-stage approach in handling high measurement error scenarios and producing more valid inference. These findings again confirm what we learned from the existing literature and emphasize the advantages of joint modeling. Overall, the above results demonstrate that our proposed model is capable of recovering the true parameter values with high coverage rates and low bias.

\begin{table}
    \begin{center}
    \caption{Simulation results including coverage rate, bias, average interval length and RMSE of $\boldsymbol{\alpha}$ comparing joint model and two-stage model approach under simulation setting 2 when the measurement error in the longitudinal marker is high.\label{tab:compare-high}} %
    \begin{tabular}{@{}llllll@{}}
    \toprule
    True Values & Model & Coverage (\%) & Average Interval Length & Bias & RMSE\\
    \midrule
    $\alpha_1$ = -0.6 & Joint Model & 94.5 & 1.28 & -0.04 & 0.34\\ 
      &  Two-Stage Model & 11.5 & 1.54 & -1.17 & 1.23\\
    $\alpha_2$ = 1 & Joint Model & 95.5 & 1.38 & 0.07 & 0.37\\
      & Two-Stage Model & 24.0 & 1.64 & 1.01 & 1.09\\
    $\alpha_3$ = -1.5 & Joint Model & 92.5 & 1.37 & -0.14 & 0.41 \\
      & Two-Stage Model & 61.0 & 1.73 & -0.73 & 0.85\\
      \bottomrule
    \end{tabular}
    \end{center}
\end{table}

\section{Application to the Michigan Work-Life Study}\label{sec2}

\subsection{Study Description}

The Michigan Work-Life Study was a randomized controlled study designed to investigate anticipatory and perseverative stress associated with potentially discriminatory and prejudicial situations \citep{sawyer2012discrimination}. 232 healthy women from a medium-sized metropolitan statistical area in Michigan, who identified as Black or African American and were between the ages of 18 and 30 years participated in a two-hour laboratory social stress test. After completing a 40-minute online survey that included sociodemographic, psychosocial, personality, and health questions, participants were randomized to one of four experimental groups for the laboratory social stress test. Throughout the entirety of the laboratory test, continuous heart rate and periodic blood pressure and stress perception data and saliva samples were collected. After exclusion of participants with substantial missing continuous heart rate data, our final analytic sample was 165. We focus our analyses on heart rate as a well-studied marker of the stress response system that is typically analyzed using information averaged over periods of time. This study was approved by the University of Michigan Institutional Review Board (HUM00134109). 

The laboratory social stress test was designed around the Trier Social Stress Test (TSST), a well-documented and well-validated mode of eliciting a biological and psychological stress response \citep{kirschbaum1993trier}. In the TSST, a participant is asked to deliver a five-minute talk about some aspect of themselves and a five-minute math task in front of an audience. The audience members are generally instructed to sit and listen to the talk and math task without facial or bodily expression. (Studies have shown that when audience members nod their heads, smile, or provide other positive cues, the TSST does not elicit the stress response from the participant \citep{wiemers2013friendly}.) In this specific study, there were three experimental groups and one control group. The three experimental groups differed in the racial composition and the racial attitudes of the audience. In Groups 1 and 2, the audience was composed of four White women and one White woman study staff member. In Group 3, the audience was composed of four Black women and one Black woman study staff member. The control group (Group 4) did not include an audience to determine whether the audience-based TSST elicited an appropriate stress response, but included a Black woman staff member. In Group 1, the participants were led to believe that the audience members were part of a human resources/administrative organization that held racially-hostile views. In Groups 2 and 3, the participants were led to believe that the audience members were part of an organization that held racially-inclusive views. Table \ref{tab:group_description} shows a detailed group specification.

\begin{table}
    \begin{center}
    \caption{Group specification. Group 1 to Group 3 are the three experimental groups and Group 4 is the control group. \textbf{WH} stands for ``White Hostile", \textbf{WF} stands for ``White Friendly", \textbf{BF} stands for ``Black Friendly" and \textbf{C} stands for ``Control".\label{tab:group_description}}
    \begin{tabular}{@{}ll@{}}
    \toprule
     Group & Audience \\
     \midrule
    Group 1 (\textbf{WH}) & Four White women and one White woman study staff member with racially-hostile views \\
    Group 2 (\textbf{WF}) & Four White women and one White woman study staff member with racially-inclusive views \\
    Group 3 (\textbf{BF}) & Four Black women and one Black woman study staff member with racially-inclusive views \\
    Group 4 (\textbf{C}) & Control group including a Black woman study staff member \\
    \bottomrule
    \end{tabular}
    \end{center}
\end{table}

\color{black} In studies using the TSST, participants have been exposed to the audience in different ways. Some studies convene a small, live audience for the participant. The limitations to this approach include the lab scheduling for hundreds of participants and, perhaps more importantly, the risk of unintentional positive cues from the audience. Others present an opaque panel and inform the participant that an audience is hidden behind the panel and in reality, there is no audience. The limitation to this approach is that it may not elicit the stress response because the facial and bodily cues are not visible to the participant. Therefore, this study simulated a virtual conference call on a large-screen television. The audience included a recording of four women appearing to listen to a talk. The advantages to this approach are that facial and bodily cues were well-controlled and that, within their experimental groups, all study participants were exposed to the exact same women and facial and bodily cues. In fact, those in Groups 1 and 2 were exposed to the exact same audience, but were led to believe that they held different racial attitudes.

Upon the arrival to the laboratory, participants completed the informed consent process. Notably, participants were led to believe that this study was focused on the ways in which potential employers view different job applicants. Only at the end of the study, during a full debriefing, were participants informed of the true research objective of the study. After obtaining informed consent, staff connected participants to an Omron blood pressure cuff (Omron Healthcare, Kyoto, Japan) and an Empatica E4 wristband (Empatica, Cambridge, MA).
% It should be noted that, with respect to the continuous heart rate measurement used for this current statistical application, these five-minute MBs are an ambiguous period.  More specifically, during the MB after the anticipation period when participants were informed that they were going to be giving a talk in front of an audience, some may have still been thinking while others may have been focused on their blood pressure and saliva collection, which would likely be reflected in their heart rate data. Further, the recovery period, which with respect to continuous heart rate, would ideally one continuous period without MBs, was broken up into five recovery periods because the blood pressure could not be measured continuously and the salivary samples were collected for the analysis of two stress biomarkers that have different lagged response times. 

The entire social stress test period is illustrated in Figure \ref{fig:exp_period}. The TSST proceeded in timed sections as follows, which are separated by delineations we call ``Marks". In the first ten-minute period (ANTICIPATION, bracketed by Mark 1 and Mark 2), the staff member explained that the participant would be giving a talk to a virtual audience of members from the (fictional) ``Michigan Human Resources Administrators Organization" and in this five-minute talk, they were to discuss why they were the ideal candidate for their dream job, whatever that job may be. The staff then walked them through a (fictional) recent survey completed by the members of this organization. The survey results showed that the organization was either composed of almost entirely White members (Groups 1 and 2) or very diverse with respect to racial composition (Group 3). Further, the survey results showed that members supported racially-hostile views (Group 1) or racially-inclusive views (Groups 2 and 3). In the remaining minutes of this period, participants could prepare for their talk with notes, but were instructed that they could not use these notes during their talk. During the ten-minute STRESS period, the staff revealed a large screen television with a supposed live audience virtual meeting of four women. During the first five minutes, bracketed by Mark 3 and Mark 4, the staff member instructed the participant to explain to the audience why they are the ideal candidate for their dream job. If they finished speaking in less than five minutes, they were instructed to continue speaking. During the second five minutes, bracketed by Mark 4 and Mark 5, the staff member instructed the participant to sequentially subtract 13, starting from 1022. If they made a mistake, they were instructed to begin again. After this stress period, the staff member turned off the television and instructed the participant to sit quietly for the remaining roughly 90 minutes (RECOVERY, bracketed by Mark 5 through Mark 11).  We limit our analysis to heart rate measurements between Mark 1 and Mark 6, to focus on the anticipation and stress periods.

In between the ANTICIPATION period and the first RECOVERY period, the Empatica E4 collected continuous heart rate information; at specific time points (called a `Measure Break' or `MB'), staff collected blood pressure, stress perception, and saliva. For a variety of reasons, these MB periods may have provided ambiguous responses; for example, some participants may have continued to think about the previous period while others may have been sufficiently distracted by the blood pressure and saliva collection so that they are not thinking about the previous period. Hence, we do include these MB periods in our analysis; however, these MB periods are not the focus of our application.

\tikzstyle{tag1} = [rectangle, 
rounded corners,
minimum width=1.5cm, 
minimum height=1cm, 
text centered, 
draw=black]

\tikzstyle{tag2} = [rectangle,  
rounded corners,
minimum width=1.5cm, 
minimum height=1cm, 
text centered, 
draw=black]

\tikzstyle{tag3} = [rectangle,  
rounded corners,
minimum width=1.5cm, 
minimum height=1cm, 
text centered, 
draw=black]

\tikzstyle{tag4} = [rectangle,  
rounded corners,
minimum width=1.5cm, 
minimum height=1cm, 
text centered, 
draw=black]

\tikzstyle{tag5} = [rectangle,  
rounded corners,
minimum width=1.5cm, 
minimum height=1cm, 
text centered, 
draw=black]

\tikzstyle{tag6} = [rectangle,  
rounded corners,
minimum width=1.5cm, 
minimum height=1cm, 
text centered, 
draw=black]

\tikzstyle{tag7} = [rectangle,  
rounded corners,
minimum width=1.5cm, 
minimum height=1cm, 
text centered, 
draw=black]

\tikzstyle{tag8} = [rectangle,  
rounded corners,
minimum width=1.5cm, 
minimum height=1cm, 
text centered, 
draw=black]

\tikzstyle{tag9} = [rectangle,  
rounded corners,
minimum width=1.5cm, 
minimum height=1cm, 
text centered, 
draw=black]

\tikzstyle{tag10} = [rectangle,  
rounded corners,
minimum width=1.5cm, 
minimum height=1cm, 
text centered, 
draw=black]

\tikzstyle{end} = [rectangle,  
rounded corners,
minimum width=1.5cm, 
minimum height=1cm, 
text centered, 
draw=black]

\tikzstyle{arrow} = [->,>=stealth]

\begin{figure}
    \centering
    \begin{tikzpicture}[node distance=3cm, x=1cm, y=1cm]
    %\draw[help lines,yellow](-7,0) grid (9,-10);

\node (tag1) [tag1] {Mark 1};
\node (tag2) [tag2, right of=tag1, xshift=3cm] {Mark 2};

\node (tag3) [tag3, below of=tag2, yshift=1cm] {Mark 3};
\node (tag4) [tag4, left of=tag3, xshift=-3cm] {Mark 4};
\node (tag5) [tag5, left of=tag4, xshift=-3cm] {Mark 5};

\node (tag6) [tag6, below of=tag4, xshift=0cm, yshift=1cm] {Mark 6};
\node (tag7) [tag7, right of=tag6, xshift=3cm] {Mark 7};

\node (tag8) [tag8, left of=tag7, xshift=-3cm, yshift=-2cm] {Mark 8};
\node (tag9) [tag9, left of=tag8, xshift=-3cm] {Mark 9};

\node (tag10) [tag10, below of=tag6, yshift=-1cm] {Mark 10};
\node (end) [end, right of=tag10] {Mark11};

\draw [arrow] (tag1) -- node[anchor=south, distance = 5] {ANTICIPATION (\textbf{p})} node[anchor=north] {10 min} (tag2);
\draw [arrow, darkgray] (tag2) -- node[anchor=east] {MB 1 (\textbf{m$_1$})} node[anchor=west] {5 min} (tag3);

\draw [arrow] (tag3) -- node[anchor=south] {STRESS 1 (SPEECH) (\textbf{s})} node[anchor=north] {5 min} (tag4);
\draw [arrow] (tag4) -- node[anchor=south] {STRESS 2 (MATH) (\textbf{s})} node[anchor=north] {5 min}(tag5);

\draw [arrow, darkgray] (-6,-2.5)  -- node[anchor=east] {MB 2 (\textbf{m$_2$})} node[anchor=west] {5 min} (-6,-4);

\draw [arrow] (-6,-4) -- node[anchor=south] {RECOVERY (\textbf{r})} node[anchor=north] {5 min} (tag6);
\draw [arrow, darkgray] (tag6) -- node[anchor=south] {MB 3} node[anchor=north] {5 min} (2,-4);
\draw [arrow] (2,-4) -- node[anchor=south] {RECOVERY} node[anchor=north] {5 min} (tag7);

\draw [arrow, darkgray] (tag7) -- node[anchor=east] {MB 4} node[anchor=west] {5 min} (6,-6);

\draw [arrow] (6,-6) -- node[anchor=south] {RECOVERY} node[anchor=north] {15 min} (tag8);
\draw [arrow, darkgray] (tag8) -- node[anchor=south] {MB 5} node[anchor=north] {5 min} (-2,-6);
\draw [arrow] (-2,-6) -- node[anchor=south] {RECOVERY} node[anchor=north] {15 min} (tag9);

\draw [arrow, darkgray] (tag9) -- node[anchor=east] {MB 6} node[anchor=west] {5 min} (-6, -8);

\draw [arrow] (-6, -8) -- node[anchor=south] {RECOVERY} node[anchor=north] {15 min} (tag10);
\draw [arrow, darkgray] (tag10) --node[anchor=south] {MB 7} node[anchor=north] {5 min} (end);
\end{tikzpicture}
    \caption{Flowchart of the entire experimental period. The focus of the statistical analysis is on the experimental period from Mark 1 to Mark 6. We labeled the five periods from Mark 1 to Mark 6 by \textbf{p}, \textbf{m$_1$}, \textbf{s}, \textbf{m$_2$} and \textbf{r} respectively.}

    \label{fig:exp_period}
\end{figure}
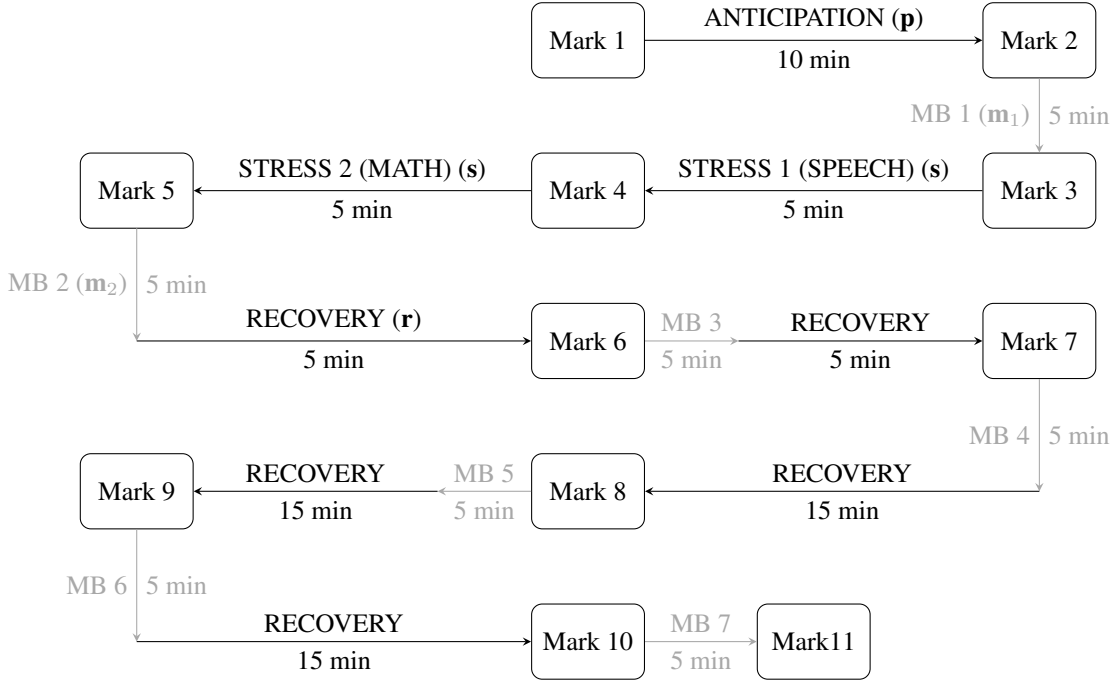

\subsection{Model Framework}\label{subsec1}

\subsubsection{Longitudinal marker Sub-model}\label{subsubsec1}

\begin{figure}[t]
    \centering
    \includegraphics[scale=0.4]{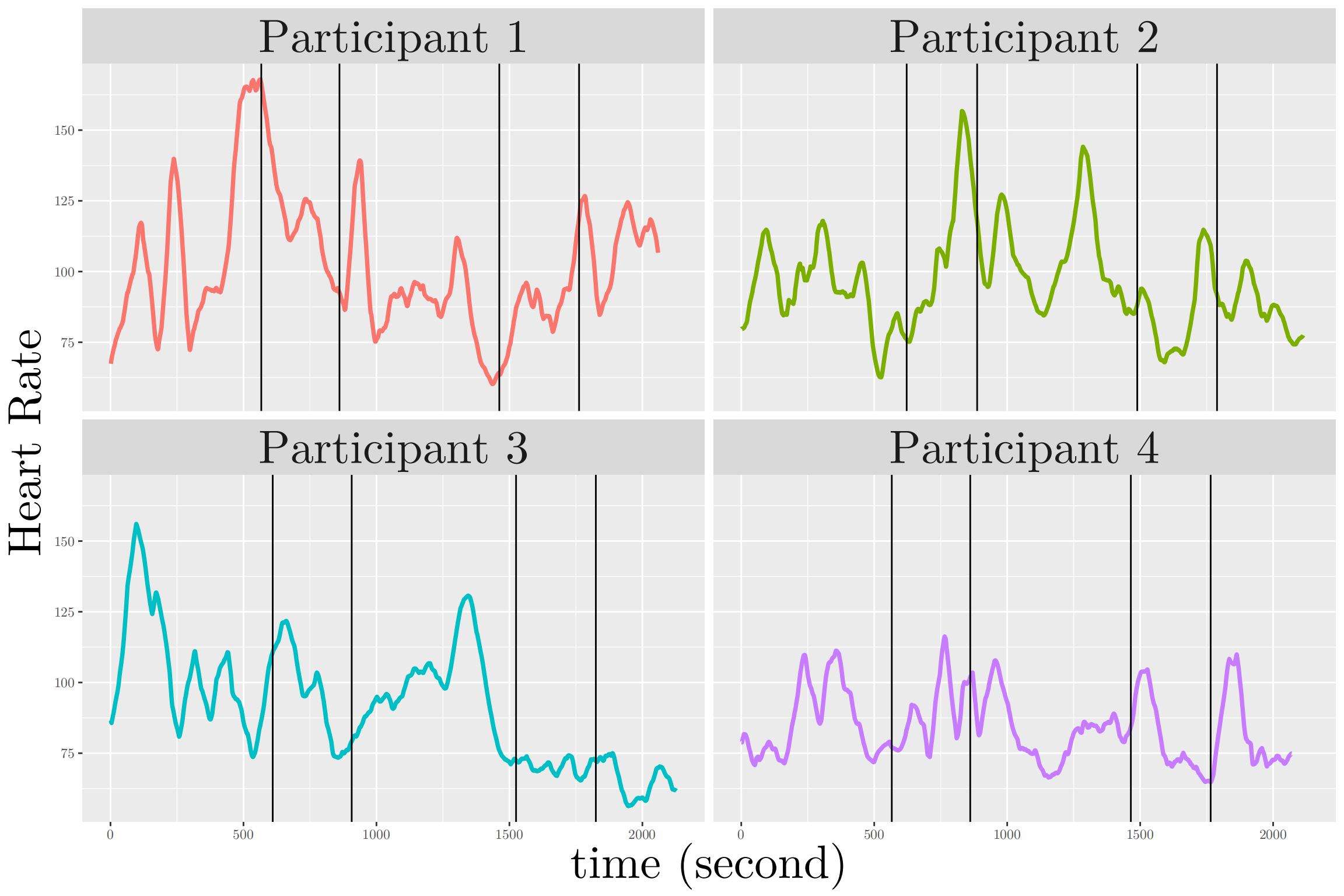}
    \caption{Heart rate measurements from Mark 1 to Mark 6 for four randomly selected participants. The four solid lines correspond to Mark 2 representing the end of the anticipation period and the start of the first measure break, Mark 3 representing the start of the stress period, Mark 4 representing the end of the stress period and the start of the second measure break, and Mark 5 representing the start of the recovery period, respectively.}
    \label{fig:sampleHR}
\end{figure}

The longitudinal marker of interest in this application is heart rate over time, which was collected each second continuously throughout the entire experimental period. 
There were an average of 2153 heart rate observations per woman between Mark 1 to Mark 6; the number of observations varied from a minimum of 1878 to a maximum of 3872. Figure \ref{fig:sampleHR} shows the heart rate measurements from Mark 1 to Mark 6 for four randomly selected participants.

We fit the cubic B-spline model with AR1 as shown in the method section. We selected the time knots $k_l$'s to be every 2nd percentiles of the observation time $t_{ij}$. However, we assumed that the long-term variability and the short-term variability differ by experimental periods (e.g., anticipation, stress) rather than the uniform long-term and short-term variability in (1) and (6). We included five periods bracketed by Mark 1 through Mark 6 (see Figure \ref{fig:exp_period}): the ten-minute anticipation period, denoted by \textbf{p}; the first (nuisance) five-minute measurement break, denoted by \textbf{m$_1$}; the ten-minute stress period, denoted by \textbf{s}; the second (nuisance) five-minute measurement break, denoted by \textbf{m$_2$}; and a five-minute recovery period, denoted by \textbf{r}. Our long-term variability model is given by
\begin{align}
    & b_{il}\overset{\mathrm{i.i.d}}\sim N(0, \sigma_{b_i}^{(\boldsymbol{p})2}\times I(\text{period}_{il}=\boldsymbol{p})+\sigma_{b_i}^{(\boldsymbol{m_1})2}\times I(\text{period}_{il}=\boldsymbol{m_1})+ \nonumber \\
            & \sigma_{b_i}^{(\boldsymbol{s})2}\times I(\text{period}_{il}=\boldsymbol{s})+\sigma_{b_i}^{(\boldsymbol{m_2})2}\times I(\text{period}_{il}=\boldsymbol{m_2})+\sigma_{b_i}^{(\boldsymbol{r})2}\times I(\text{period}_{il}=\boldsymbol{r})),
            l=1,...,L,
\end{align}
where $I(\cdot)$ is an indicator function indicating whether the $l$-th random effect of the $i$-th participant falls into a specific experimental period. We also assumed that the long-term variability in these different periods had their own hyperpriors:
\begin{align}
            & \log(\sigma_{b_i}^{(\boldsymbol{p})2})\sim N(v_{b}^{(\boldsymbol{p})},\Psi_{b}^{(\boldsymbol{p})2}), \ \ \log(\sigma_{b_i}^{(\boldsymbol{m_1})2})\sim N(v_{b}^{(\boldsymbol{m_1})}, \Psi_{b}^{(\boldsymbol{m_1})2}), \nonumber \\
            & \log(\sigma_{b_i}^{(\boldsymbol{s})2})\sim N(v_{b}^{(\boldsymbol{s})},\Psi_{b}^{(\boldsymbol{s})2}), \ \ \log(\sigma_{b_i}^{(\boldsymbol{m_2})2})\sim N(v_{b}^{(\boldsymbol{m_2})}, \Psi_{b}^{(\boldsymbol{m_2})2}), \ \ 
            \log(\sigma_{b_i}^{(\boldsymbol{r})2})\sim N(v_{b}^{(\boldsymbol{r})},\Psi_{b}^{(\boldsymbol{r})2}).
\end{align}

Similarly, we extended (6) to 
\begin{align}
    & w_{ij}\overset{\mathrm{i.i.d}}\sim N(0, \sigma_{w_i}^{(\boldsymbol{p})2}\times I(\text{period}_{ij}=\boldsymbol{p})+\sigma_{w_i}^{(\boldsymbol{m_1})2}\times I(\text{period}_{ij}=\boldsymbol{m_1})+ \nonumber \\
            & \sigma_{w_i}^{(\boldsymbol{s})2}\times I(\text{period}_{ij}=\boldsymbol{s})+\sigma_{w_i}^{(\boldsymbol{m_2})2}\times I(\text{period}_{ij}=\boldsymbol{m_2})+\sigma_{w_i}^{(\boldsymbol{r})2}\times I(\text{period}_{ij}=\boldsymbol{r})),
            j=1,...,n_i.
\end{align}

Similarly, as the long-term variability, we assumed that the short-term variability in these different periods had their own hyperpriors: 
\begin{align}
            & \log(\sigma_{w_i}^{(\boldsymbol{p})2})\sim N(v_{w}^{(\boldsymbol{p})},\Psi_{w}^{(\boldsymbol{p})2}), \ \ \log(\sigma_{w_i}^{(\boldsymbol{m_1})2})\sim N(v_w^{(\boldsymbol{m_1})}, \Psi_{w}^{(\boldsymbol{m_1})2}), \nonumber \\            & \log(\sigma_{w_i}^{(\boldsymbol{s})2})\sim N(v_{w}^{(\boldsymbol{s})},\Psi_{w}^{(\boldsymbol{s})2}), \ \ \log(\sigma_{w_i}^{(\boldsymbol{m_2})2})\sim N(v_{w}^{(\boldsymbol{m_2})}, \Psi_{w}^{(\boldsymbol{m_2})2}), \ \ 
            \log(\sigma_{w_i}^{(\boldsymbol{r})2})\sim N(v_{w}^{(\boldsymbol{r})},\Psi_{w}^{(\boldsymbol{r})2}).
\end{align}

\begin{figure}[t]
    \centering
    \includegraphics[scale=0.55]{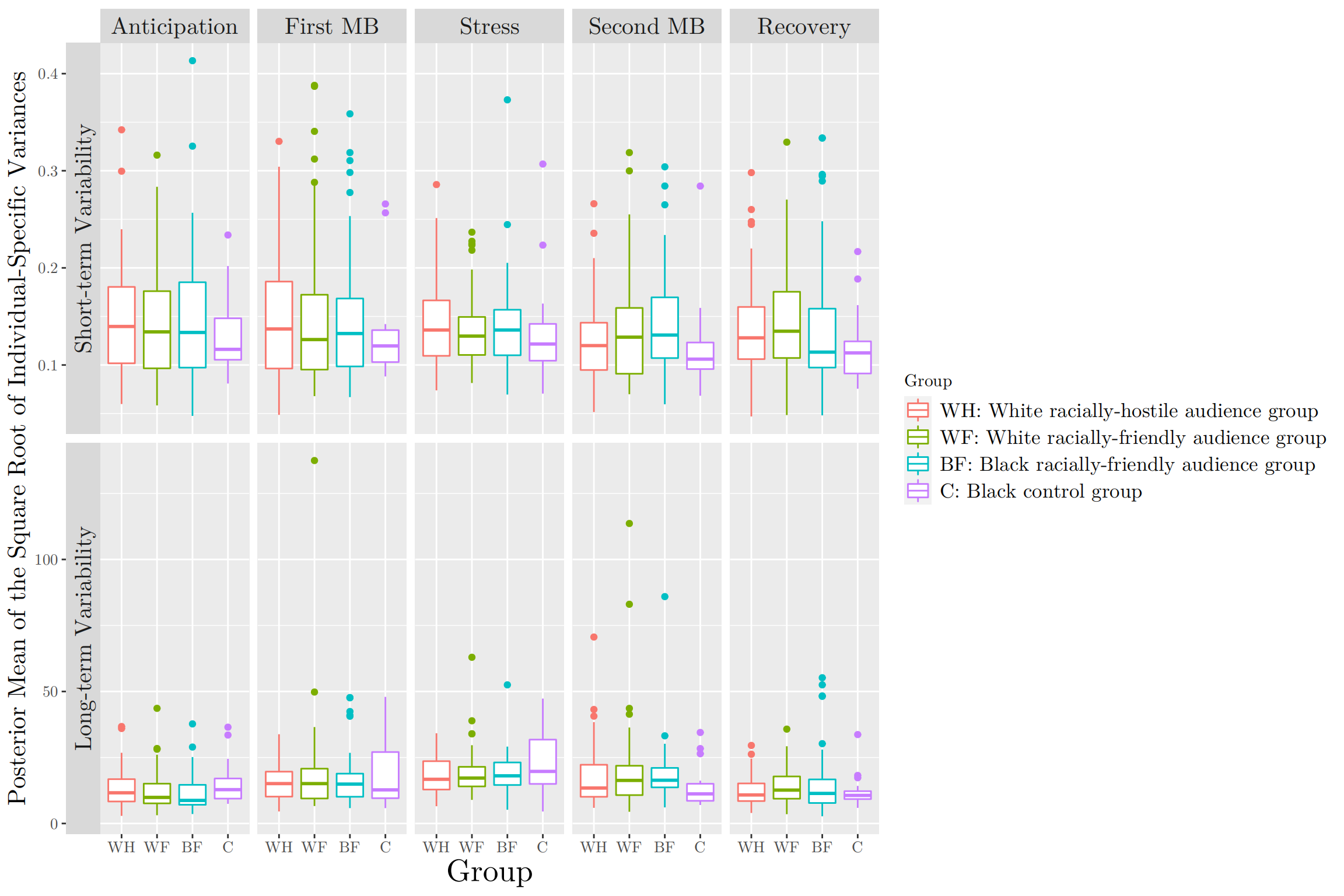}
    \caption{Short- and long-term variability in each of the experimental periods across groups characterized by the posterior means of the square root of individual-specific variances. The top five panels correspond to the posterior means of $\sigma_{w_i}^{(\boldsymbol{p})}$'s, $\sigma_{w_i}^{(\boldsymbol{m_1})}$'s, $\sigma_{w_i}^{(\boldsymbol{s})}$'s, $\sigma_{w_i}^{(\boldsymbol{m_2})}$'s and $\sigma_{w_i}^{(\boldsymbol{r})}$'s respectively and the bottom five panels correspond to the posterior means of $\sigma_{b_i}^{(\boldsymbol{p})}$'s, $\sigma_{b_i}^{(\boldsymbol{m_2})}$'s, $\sigma_{b_i}^{(\boldsymbol{s})}$'s, $\sigma_{b_i}^{(\boldsymbol{m_2})}$'s and $\sigma_{b_i}^{(\boldsymbol{r})}$'s respectively}
    \label{fig:posterior_var}
\end{figure}

The posterior means of the square root of individual-level variances, $\sigma_{w_i}^{(\cdot)}$'s and $
\sigma_{b_i}^{(\cdot)}$'s, across the four experimental groups in each of the experimental period is shown in Figure \ref{fig:posterior_var}.

\subsubsection{Outcome Sub-model}

The outcome of interest in this application is the binary group assignment: Group 1 vs Group 2 with Group 1 as the reference group, Group 2 vs Group 3 with Group 2 as the reference group, Group 1 vs Group 4 with Group 4 as the reference group, Group 2 vs Group 4 with Group 4 as the reference group, or Group 3 vs Group 4 with Group 4 as the reference group. Pairwise group comparisons follow the substantive meaning represented by the experimental conditions. Comparing Groups 1 and 2 allows us to focus on potential variation introduced by the racial attitudes of the audience (hostile or friendly) because the racial composition of the audience was the same between groups. Comparing Groups 2 and 3 allows us to focus on the racial composition (Black or White) because the racial attitudes of the audience were the same. We also compared each group with the control group in order to examine whether there was an impact of public speaking (the core of the TSST). Note that we used group assignment as our outcome mainly to study its association with short-term/long-term variability in heart rate, which is thought to reflect the anticipation and perseveration about potential discriminatory and prejudicial situations reflected by the group assignment, rather than testing the causal relationship. Since we had different variances in different experimental periods now, we extended the linear prediction function in (2) to 
\begin{align}
    & \eta(S_i)=\alpha_1+\alpha_2\times \log(\sigma_{w_i}^{(\boldsymbol{p})}\times 10)+\alpha_3\times \log(\sigma_{w_i}^{(\boldsymbol{m_1})}\times 10)+\alpha_4\times \log(\sigma_{w_i}^{(\boldsymbol{s})}\times 10) \nonumber \\
    & +\alpha_5\times \log(\sigma_{w_i}^{(\boldsymbol{m_2})}\times 10)+
    \alpha_6\times \log(\sigma_{w_i}^{(\boldsymbol{r})}\times 10)+\alpha_7\times \log(\frac{\sigma_{b_i}^{(\boldsymbol{p})}}{10})+\alpha_8\times \log(\frac{\sigma_{b_i}^{(\boldsymbol{m_1})}}{10})+\alpha_9\times \log(\frac{\sigma_{b_i}^{(\boldsymbol{s})}}{10})
   \nonumber \\ & +\alpha_{10}\times \log(\frac{\sigma_{b_i}^{(\boldsymbol{m_2})}}{10})+\alpha_{11}\times \log(\frac{\sigma_{b_i}^{(\boldsymbol{r})}}{10}),
\end{align}
where we scaled the long-term and short-term variances inside the log function to make the chains in the posterior sampling procedure have better mixing performance. We did not include any other covariates into the outcome model since the participants were randomly assigned and their covariates values were balanced.

For the analysis, we ran 3 chains with 18,000 iterations each, treating the first 9,000 as burn-in. We again assessed the convergence of the models by Gelman-Rubin $\hat{R}$ measure, showing maximum values under 1.1 and thus good convergence.

\subsection{Results}

Drawing from the TSST and stress biology literature as well as the humanities and social science literature, we hypothesize the following patterns of results across the TSST periods and experimental groups (See Table \ref{tab:Hypothesis}). We hypothesize that, during all periods, participants exposed to the White, racially-hostile audience will experience more stress compared to those exposed to the White racially-friendly audience. We further hypothesize that, during the anticipation period, participants exposed to the Black, racially-friendly audience and the White, racially-friendly audience will experience roughly similar levels of stress. This is because they will have been told the same things about the speaking activity and the audience. The only difference between these groups is the race of the staff member. However, both the Black and the White staff members interacted in a civil but stoic manner with participants. Interacting with another Black woman who is not overtly welcoming in a new space (the lab) may be a source of stress. The social science literature discuss the importance of social cues of affirmation among Black men women, even among strangers, in public. Thus, during the stress and recovery periods, we hypothesize that participants exposed to the stoic audience of Black women, even though they were primed to believe that they were racially-friendly, will experience greater stress compared to those exposed White, racially-friendly audience. Finally, we hypothesize that participants exposed to any of the audiences will experience greater stress compared to those who were exposed only to the (Black) staff member. However, there is the possibility that those exposed to the White, friendly audience may experience lower levels of stress compared to those in the control group, because this latter group still had to interact with the Black staff member who carried herself in a stoic manner.

\begin{table}
\begin{center}
\caption{Hypothesized differences in stress response across TSST periods and the experimental groups. 
    Abbreviations: BF, Black friendly audience group; C, control group; WF, White friendly audience group; WH, White hostile audience group; ND, no difference.\label{tab:Hypothesis}} %
    \begin{tabular}{@{}llllll@{}}
    \toprule
    TSST Period & WF vs WH & BF vs WF & WH vs C & WF vs C & BF vs C \\
    \midrule
    Anticipation & WH $>$ WF & ND & WH $>$ C & WF $>$ 
    C & BF $>$ C\\
    Stress & WH $>$ WF & BF $>$ WF & WH $>$ C & WF $>$ 
    C & BF $>$ C\\
    Recovery & WH $>$ WF & BF $>$ WF & WH $>$ C & WF $>$ 
    C & BF $>$ C\\
    \bottomrule
    \end{tabular}
    \end{center}
\end{table}

The results from the application of our analytic approach suggest that there is a difference in the stress response meaning of second-to-second (short-term) and minute-to-minute (long-term) patterns in variability. Table \ref{tab:Composite} provides the results from all of the separate models together to facilitate interpretation. Note that each column represents the results from a separate model. When considering the results for second-to-second variability, participants in the control group (conceptualized as the lowest stress exposure group, as there was no exposure to any audience) did not exhibit higher variability in heart rate in any phase compared to any of the other experimental groups (Table \ref{tab:Composite}, top panel). However, when considering minute-to-minute variability, participants in the control group exhibited greater variability in heart rate compared to each of the three other exposure groups at different phases (Table \ref{tab:Composite}, bottom panel). For example, participants in the control group exhibited greater minute-to-minute variability compared to those exposed to the Black friendly audience during the anticipation and stress phases. While this pattern was not universal, as we discuss below, the difference in the pattern of results for the second-to-second and minute-to-minute variability suggests that each of these types of variability captures something different about the stress response process.

\begin{sidewaystable} 
\caption{Coefficients and 95\% credible intervals for each of the experimental group comparisons for second-to-second variability ($\alpha_2,...,\alpha_6$) and minute-to-minute variability ($\alpha_7,...,\alpha_{11}$). Abbreviations: BF, Black friendly audience group; C, control group; MB, measurement break; ref, reference group; WF, White friendly audience group; WH, White hostile audience group\label{tab:Composite}} %
\tabcolsep=0pt%
    \begin{tabular*}{\textwidth}{@{\extracolsep{\fill}}lccccc@{\extracolsep{\fill}}}
    \toprule%
    Coefficient (Period) & WF vs WH (ref) & BF vs WF (ref) & WH vs C (ref) & WF vs C (ref) & BF vs C (ref) \\
    \midrule
    $\alpha_1$ (Intercept) & -0.12 & 0.02 & 4.83 & 0.31 & 0.80\\
     & (-0.73, 0.45) & (-0.55, 0.60) & (0.69, 9.99) & (-3.44, 4.60) & (-2.83, 4.78)\\
    $\alpha_2$ (Anticipation) & 0.19 & 0.22 & 4.59 & 8.18 & \textbf{12.50}\\
     & (-1.12, 1.55) & (-0.83, 0.59) & (-6.25, 16.12) & (-3.00, 19.79) & \textbf{(2.03, 22.96)}\\
    $\alpha_3$ (MB1) & 0.16 & -0.11 & 8.67 & 2.82 & -4.23\\
     & (-0.84, 1.17) & (-0.96, 0.74) & (-0.04, 18.82) & (-4.27, 10.38) & (-12.16, 3.01)\\
    $\alpha_4$ (Stress) & -1.35 & -0.12 & 6.73 & 0.11 & 2.14\\
     & (-2.91, 0.10) & (-1.54, 1.25) & (-2.81, 17.56) & (-8.91, 9.36) & (-5.80, 10.89)\\
    $\alpha_5$ (MB2) & 0.60 & 0.64 & -2.07 & \textbf{10.87} & \textbf{8.65}\\
     & (-0.54, 1.79) & (-0.33, 1.60) & (-12.09, 7.88) & \textbf{(2.27, 20.72)} & \textbf{(1.02, 18.12)}\\  
    $\alpha_6$ (Recovery) & 0.14 & -0.71 & 5.98 & \textbf{11.42} & -3.66\\
     & (-0.90, 1.24) & (-1.83, 0.34) & (-3.28, 16.57) & \textbf{(1.58, 22.01)} & (-12.95, 5.52)\\
    $\alpha_7$ (Anticipation) & -0.72 & -0.08 & 5.88 & -4.13 & \textbf{-15.01}\\
     & (-1.87, 0.28) & (-0.86, 0.72) & (-2.15, 14.06) & (-11.80, 2.24) & \textbf{(-25.68, -2.01)}\\
    $\alpha_8$ (MB1) & 0.30 & -0.17 & \textbf{-23.62} & \textbf{-14.52} & 2.97\\
     & (-0.66, 1.37) & (-0.85, 0.48) & \textbf{(-36.20, -13.73)} & \textbf{(-23.13, -6.38)} & (-3.24, 9.41)\\
    $\alpha_9$ (Stress) & 0.53 & 0.08 & 1.91 & \textbf{12.30} & \textbf{-8.64}\\
     & (-0.43, 1.54) & (-0.82, 1.01) & (-4.65, 9.28) & \textbf{(5.69, 20.13)} & \textbf{(-17.22, -0.43)}\\
    $\alpha_{10}$ (MB2) & 0.07 & -0.05 & 4.32 & \textbf{-5.87} & \textbf{12.50}\\
     & (-0.56, 0.71) & (-0.65, 0.56) & (-1.79, 12.05) & \textbf{(-11.91, -0.79)} & \textbf{(0.90, 24.19)}\\
    $\alpha_{11}$ (Recovery) & 0.25 & 0.25 & -1.76 & -3.98 & -1.44\\
     & (-0.54, 1.07) & (-0.36, 0.94) & (-11.64, 7.32) & (-12.41, 4.75) & (-8.52, 5.58)\\
     \bottomrule
     \end{tabular*}
\end{sidewaystable}

Minute-to-minute (long-term) variability may capture what is thought to be captured by standard heart rate variability that is averaged over time. The general literature on stress and heart rate variability suggests that greater mean heart rate variability, averaged over time (e.g., a 5-minute period), is related to lower heart rate and lower acute stress exposure. Compared to participants in each of the three audience-based stress exposure groups, those in the control group (conceptualized as the least stressful exposure group) exhibited greater variability in heart rate. Specifically, participants in the control group exhibited greater variability compared to those exposed to the White hostile audience during the break period that immediately followed the anticipation period ($\alpha_8$: -23.62; 95\%CI: -36.20, -13.73); compared to those exposed to the White friendly audience during the breaks immediately following both the anticipation and stress periods ($\alpha_8$: -14.52; 95\%CI: -23.13, -6.38; $\alpha_{10}$: -5.87; 95\%CI: -11.91, -0.79); and compared to those exposed to the Black friendly audience during the anticipation and stress periods ($\alpha_7$: -15.01; 95\%CI: -25.68, -2.01; $\alpha_9$: -8.64; 95\%CI: -17.22, -0.43). This suggests that those in the control group experienced less stress exposure compared to those in the other three groups.

Notably, however, this pattern of greater minute-to-minute variability with the control group compared to other groups was not universal. If we are to understand the minute-to-minute variability as inversely related to stress response, then, during the stress period, participants exposed to the White, racially-friendly audience experienced less stress compared to those in the control group. As discussed above, it may be that interaction with a stoic, Black staff member resulted in similar stress experiences as delivering a talk in front of a stoic, White, racially-friendly audience. Our results also suggest that, during the break immediately following the stress period, participants exposed to the Black, racially-friendly audience experienced less stress compared to those in the control group. It may be that there are similar levels of stress experienced when interacting with a stoic panel of Black women virtually and with a Black staff member in person.

Our results suggest that second-to-second (short-term) variability in heart rate captures something different about the stress response process and is perhaps positively (rather than inversely) correlated with the stress experience. During the stress period, participants exposed to the White, racially-hostile audience experienced greater variability compared to those exposed to the White, racially-friendly audience ($\alpha_4$: -1.35; 95\%CI: -2.91, 0.10), although the 95\%CI included zero. Similarly, when there were differences in the variability in heart rate between any of the three audience-based groups and the control group, our results suggest that participants delivering a talk in front of an audience experienced greater stress. This pattern is consistent with that shown with minute-to-minute variability.

\subsection{Posterior Predictive Model Checking}

\begin{figure}
    \centering
    \includegraphics[scale=0.85]{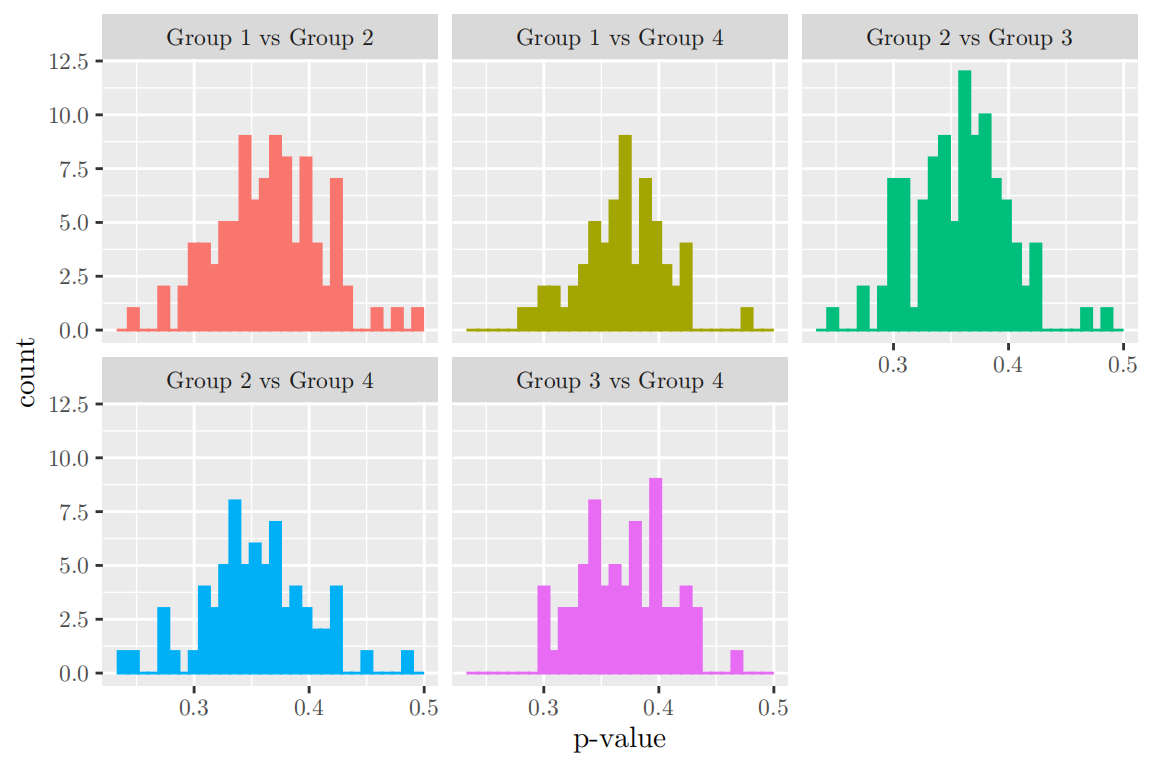}
    \caption{Posterior predictive check for longitudinal marker data across all five models in the application. Each panel corresponds to the histogram of the posterior predictive p-values across all individuals in one specific model. The top three panels correspond to the model comparing Group 1 with Group 2 (median: 0.364), the model comparing Group 1 with Group 4 (median: 0.372) and the model comparing Group 2 with Group 3 (median: 0.362), respectively; the bottom two panels correspond to the model comparing Group 2 with Group 4 (median: 0.353) and the model comparing Group 3 with Group 4 (median: 0.368), respectively.}
    \label{fig:PPD check}
\end{figure}

To assess whether the proposed model produces measurements that approximate the true data from the Michigan Work-Life Study, we further conducted a ``goodness-of-fit" test on both the longitudinal marker sub-model and the binary outcome sub-model based on the posterior predictive distribution (PPD). Specifically, we calculated the posterior predictive p-values \citep{gelman1996posterior}, which examined the probability that a test statistic calculated from the replicated data $X^{rep}$ is more extreme than the same test statistic calculated from the observed data $X$. For the longitudinal marker data, the replicated data $X^{rep}$ is drawn from the posterior predictive distribution $f(X^{rep}|X)=\int f(X^{rep}|\theta,X)p(\theta|X)d\theta$ where $X$ is the observed longitudinal marker data used to fit the model and $\theta$ is the unknown model parameters. We obtained $X^{rep}$ from the following steps, note that we included the first-order autocorrelation structure when simulating $X^{rep}$: 
\begin{enumerate}
    \item We calculated the fitted values from the mean structure of the longitudinal marker $f(t_{ij}; \beta, \boldsymbol{b}_i)=\beta_0+\sum_{l=1}^Lb_{il}\phi_{l,3}(t_{ij})$.
    \item We generated ``white noise" $w_{ij}$ from $N(0, \sigma_{w_i}^{(p)2}\times I(\text{period}_{ij}=p)+\sigma_{w_i}^{(m_1)2}\times I(\text{period}_{ij}=m_1)+\sigma_{w_i}^{(s)2}\times I(\text{period}_{ij}=s)+\sigma_{w_i}^{(m_2)2}\times I(\text{period}_{ij}=m_2)+\sigma_{w_i}^{(r)2}\times I(\text{period}_{ij}=r))$.
    \item We initiated error $e_{i0}$ by $e_{i0}\sim N(0, \frac{\sigma_{w_i}^{(p)2}}{1-\rho_i^2})$ and then generated a series of errors with AR1 correlation structure as $e_{ij}=\rho_ie_{ij-1}+w_{ij}$. 
    \item We calculated fitted values incorporating AR1 correlation structure by $\Tilde{f}(t_{ij};\beta,\boldsymbol{b}_i)=f(t_{ij};\beta, \boldsymbol{b}_i) + \rho_ie_{ij-1}$ with $w_{ij}$ left over as random errors.
    \item We drew $X^{rep}$ from the posterior predictive distribution as $X^{rep}_{ij}=f(t_{ij};\beta, \boldsymbol{b}_i)+e_{ij}$.
\end{enumerate}
    Then, we defined a chi-square discrepancy statistics as $T_i(\boldsymbol{X}_i; \beta, \boldsymbol{b}_i,\boldsymbol{\sigma}^2_i)=\sum_{j=1}^{n_i}\frac{(X_{ij}-\Tilde{f}(t_{ij};\beta, \boldsymbol{b}_i))^2}
{\sigma^2_i}$, where
$\sigma^2_i=\sigma_{w_i}^{(p)2}\times I(\text{period}_{ij}=p)+\sigma_{w_i}^{(m_1)2}\times I(\text{period}_{ij}=m_1)+\sigma_{w_i}^{(s)2}\times I(\text{period}_{ij}=s)+\sigma_{w_i}^{(m_2)2}\times I(\text{period}_{ij}=m_2)+\sigma_{w_i}^{(r)2}\times I(\text{period}_{ij}=r)$. We computed 500 values of $T_i(\boldsymbol{X}_i^{obs};\beta, \boldsymbol{b}_i,\boldsymbol{\sigma}^2_i)$ from both the observed longitudinal marker data and 500 draws from the posterior of $\beta$, $\boldsymbol{b}_i$ and $\boldsymbol{\sigma}_i^2$ and 500 values of $T_i(\boldsymbol{X}_i^{rep};\beta, \boldsymbol{b}_i,\boldsymbol{\sigma}^2_i)$. We compared the test statistics obtained from the observed data and the replicated data by the posterior predictive p-value defined as $P(T_i(\boldsymbol{X}_i^{obs};\beta, \boldsymbol{b}_i,\boldsymbol{\sigma}^2_i)<T_i(\boldsymbol{X}_i^{rep};\beta, \boldsymbol{b}_i,\boldsymbol{\sigma}^2_i)|(\boldsymbol{X}_i^{obs}))$. Since we fit five joint models in the actual application, we conducted the posterior predictive check on the longitudinal marker in all five models. Figure \ref{fig:PPD check} displays the histograms of the posterior predictive p-values in each of the five models. Most of the computed p-values fall between 0.2 and 0.5 and the median values of the posterior predictive p-values in these models are all around 0.36, which indicates that the replicated values from the longitudinal marker sub-model are relatively close to the observed values used to fit the model and the longitudinal sub-model fits the observed marker data well.

We further checked the posterior predictive distribution of the binary outcome $Y_i$. For the binary outcome data, the replicated data $Y_i^{rep}$ is drawn from a Bernoulli distribution with probability $\pi_i=\Phi(\eta(\boldsymbol{\sigma}_i, \boldsymbol{\alpha}))$. We considered the test statistic as the total number of events $T(\boldsymbol{Y})=\sum_{i=1}^NY_i$ and we computed 500 values of $T(\boldsymbol{Y}^{rep})$ from 500 draws from the posterior of $\boldsymbol{\sigma}_i$ and $\boldsymbol{\alpha}$. We compared the test statistics obtained from the replicated data with the total number of events in the observed binary outcomes by the posterior predictive p-values $P(T(\boldsymbol{Y}^{obs})<T(\boldsymbol{Y}^{rep})|\boldsymbol{Y}^{obs})$. Similarly to what we did with checking the longitudinal marker sub-model, we repeated the posterior predictive check on the binary outcomes in all five models. Table \ref{tab:PPD_outcome} shows the total number of events in the observed binary outcomes $T(\boldsymbol{Y}^{obs})$ and the posterior predictive p-value in each of the five models. The p-values are all around 0.4, which means that the replicated binary outcomes from the outcome sub-model are close to the observed data and the outcome sub-model is a good fit of the observed data.

\begin{table}[t]
    \begin{center}
    \caption{Posterior predictive check for the binary outcomes across all five models in the application. $T(\boldsymbol{Y}^{obs})$ represents the total number of events in the observed binary outcomes from each specific model.\label{tab:PPD_outcome}} %
    \begin{tabular}{@{}lll@{}}
    \toprule
     & $T(\boldsymbol{Y}^{obs})$ & p-value \\
     \midrule
     Group 1 vs Group 2 & 52 & 0.45 \\
     Group 1 vs Group 4 & 43 & 0.36 \\
     Group 2 vs Group 3 & 52 & 0.46 \\
     Group 2 vs Group 4 & 52 & 0.38 \\
     Group 3 vs Group 4 & 52 & 0.34 \\
     \bottomrule
    \end{tabular}
    \end{center}
\end{table}

\section{Discussion}

This paper presents a joint modeling approach that connects variabilities in the longitudinal marker data with a binary outcome. Through simulation studies, we have demonstrated that our joint model performs better than the commonly used two-stage approach when relating longitudinal measurements with a cross-sectional outcome. We applied our method to the Michigan Work-Life Study and investigated the association between social stress and variabilities in the longitudinal measurements of heart rate. We found that participants exposed to a higher vigilance situation typically exhibited a significantly higher short-term variability and lower long-term variability when they were asked to prepare for or perform the Trier Social Stress Test (TSST).

Our work is important in terms of both methodological development of joint modeling and application of heart rate data. Joint modeling proves to be superior to the two-stage approach in that joint modeling yields much more accurate estimates. Unlike the two-stage approach, where uncertainty from the first stage cannot be carried over to the second stage, joint modeling allows for the propagation of the uncertainty from the longitudinal marker sub-model to the outcome sub-model. As shown in the simulation studies, our proposed joint model consistently achieves high coverage rates and low bias for the unknown parameters whereas the two-stage approach performed poorly when the measurement error in the longitudinal marker data increases. Our work extended the existing joint modeling approaches to intensive longitudinal measurements setting which consists of thousands of measurements and incorporated both the residual and random effects variabilities as predictors in the outcome sub-model, which is a novel contribution compared to the previous joint modeling study on variabilities as predictors which focused primarily on the residual variances. We termed the variability of the random effects as ``long-term" variability and the variability of the residuals as ``short-term" variability, which characterized individual-level variabilities into two types. This approach of joint modeling is rare and could be useful in modeling intensive longitudinal marker data, especially when the number of random effects in the modeling is high and individual interpretation of each random effect becomes challenging. Our work is also innovative in terms of the application of heart rate data. The traditional approach of dealing with longitudinal measurements of heart rate in existing studies is through heart rate variability (HRV), where the time and frequency domain of HRV are first calculated and then used in subsequent analysis. In contrast, our model is able to jointly model longitudinal heart rate and outcomes as well as connecting different levels of variabilities in heart rate with health outcomes.

We acknowledge certain limitations in our work. First of all, we suffered from a small sample size when applying our proposed model to the actual data. There were only 18 participants in the control group and approximately 50 participants in each experimental group after removing participants with a significant amount of missing heart rate data, which makes models comparing each experimental group with the control group converge in a much longer time, as we were using a small number of data points to estimate eleven coefficients. 

Several extensions of this work can be made in the future. One extension is to include additional biomarkers in the analysis. For example, in the Michigan Work-Life Study, the Empatica E4 wristband not only collects continuous heart rate measurements but also monitors continuous electrodermal activity (EDA) measurements and skin temperature. Introducing more biomarkers into the analysis can help us find out how variabilities in different biological signals are connected with stress. However, this extension requires modeling the correlation between biomarkers, demanding more efficient modeling techniques given that modeling one single intensive longitudinal marker is already time-consuming. Another potential extension of the current work could be to allow individual-specific variances to vary by time. In our application, we implemented a simple version of this concept, where we allowed individual-specific variances to differ across experimental periods. A more sophisticated approach, however, is to model $\sigma_{e_i}$'s as a function of time $t_{ij}$. This extension may better capture the changes in biomarkers since such data like heart rate are highly variable.

%%\bibliographystyle{unsrtnat}
%%\bibliography{references}  %%% Uncomment this line and comment out the ``thebibliography'' section below to use the external .bib file (using bibtex) .





% \end{thebibliography}

\end{document}